\documentclass[journal]{IEEEtran}
\usepackage{amsmath}
\usepackage{graphicx}
\usepackage{amsfonts} 
\usepackage{amssymb}
\usepackage{amsthm}
\usepackage{cite}
\usepackage{array}

\usepackage{algorithm,algorithmic}
\makeatletter
\newcommand*{\rom}[1]{\expandafter\@slowromancap\romannumeral #1@}
\makeatother

\ifCLASSINFOpdf
\else
\fi
\begin{document}
%
\title{Joint Smoothing, Tracking, and Forecasting Based on Continuous-Time Target Trajectory Fitting}
%
%
%

\author{Tiancheng~Li,
        ~Huimin Chen, 
        ~Shudong Sun, 
        and ~Juan M. Corchado
\thanks{T. Li and J.M. Corchado are with School of Sciences, University of Salamanca, 37007 Salamanca, Spain, E-mail: \{t.c.li, corchado\}@usal.es}
\thanks{T. Li is presently visiting the Institute of Telecommunications, Vienna University of Technology, 1040 Wien, Austria}
\thanks{H. Chen is with Department of Electrical Engineering, University of New Orleans, LA 70148, USA, E-mail: hchen2@uno.edu}
\thanks{S. Sun is with School of Mechanical Engineering, Northwestern Polytechnical University, Xi'an 710072, China, E-mail: sdsun@nwpu.edu.cn}
\thanks{This work is in part supported by the Marie Sk\l{}odowska-Curie Individual Fellowship (H2020-MSCA-IF-2015) under Grant no. 709267.
}}

%
%

\markboth{T. Li, et al. Continuous-Time Target Trajectory Fitting}%
{Li \MakeLowercase{\textit{et al.}}: Continuous-Time Trajectory Fitting}
%



\maketitle

\begin{abstract}
We present a continuous time state estimation framework that unifies traditionally individual tasks of smoothing, tracking, and forecasting (STF), for a class of targets subject to smooth motion processes, e.g., the target moves with nearly constant acceleration or affected by insignificant noises. Fundamentally different from the conventional Markov transition formulation, the state process is modeled by a continuous trajectory function of time (FoT) and the STF problem is formulated as an online data fitting problem with the goal of finding the trajectory FoT that best fits the observations in a sliding time-window. Then, the state of the target, whether the past (namely, smoothing), the current (filtering) or the near-future (forecasting), can be inferred from the FoT. Our framework releases stringent statistical modeling of the target motion in real time, and is applicable to a broad range of real world targets of significance such as passenger aircraft and ships which move on scheduled, (segmented) smooth paths but little statistical knowledge is given about their real time movement and even about the sensors. 
In addition, the proposed STF framework inherits the advantages of data fitting for accommodating arbitrary sensor revisit time, target maneuvering and missed detection. The proposed method is compared with state of the art estimators in scenarios of either maneuvering or non-maneuvering target.
\end{abstract}


%
\IEEEpeerreviewmaketitle

\section{Introduction}
%
%
%
%
\IEEEPARstart{D}{ynamic} state estimation, e.g., tracking the movement of an aircraft or a car, has been widely required in engineering, which basically concerns inferring the latent state of interest based on discrete time series noisy 
observations\cite{haykin14,Sayed08,Bar-Shalom01}. The time of interest may be the past (usually referred to as smoothing), the present (filtering or tracking) or the future (prediction or forecasting). In this paper, we present an estimation framework that unifies the tasks of smoothing, tracking, and forecasting (STF), and accommodates the scenario of little a prior statistical information about the target and imperfect sensors that possibly suffer from unknown noise statistics and missed detections. For this challenging goal, this paper focuses on a specific class of targets that move in smooth patterns (e.g., moves in predefined runways and/or with nearly constant velocity/acceleration). While we do not make a rigorous definition of the ``\textit{smoothness}", we note that it corresponds to an important class of real world targets involved in air/maritime/space traffic management where for the passengers' safety, no abrupt and significant changes should be made on the movement of the carrier. 

The remainder of this paper is organized as follows. Section \ref{motivation} outlines the motivation and key contribution of our work. Two major challenges to the popular HMM (hidden Markov model) framework are also discussed. 
Section \ref{fot} presents our proposal for modeling the target motion by a deterministic trajectory function of time (FoT) and details how to obtain it from the online sensor data series. Section \ref{stf} addresses joint STF based on the trajectory FoT. Section \ref{relatedwork} reviews related works on trajectory estimation and sensor data fitting in wide disciplines, highlighting the innovation of our approach
. Section \ref{simulation} provides simulation studies to demonstrate the effectiveness of the proposed STF framework on a variety of typical scenarios with comparison to state-of-the-art smoothers and filters. Section \ref{conclusion} concludes the paper. 

\section{Motivation and Key Contribution} \label{motivation}

Generally speaking, system modeling is the prerequisite for estimation, which describes how we understand the system and the observation data. The goal of modeling is twofold: one is to relate the sensor observation to the latent target state by an observation function, and the other is to relate the target state to the time by a state function. The former explains how the data are generated from the target state, and the latter explains how the target state evolves over time. Usually, the statistical property of the sensor is easy to be estimated or given a priori and so can be considered time-invariant while that of the latent target is unknown,  complicated and time-varying, offering no real time information to the tracker. 

Due to the repetitively revisit nature of sensors, the observation function 
is commonly formulated in discrete time series as
\begin{equation} \label{eq:1}
\mathbf{y}_k = h_k(\mathbf{x}_k,\mathbf{v}_k),
\end{equation}
where $k\in \mathbb{N}$ indicates the time-instant, $\mathbf{x}_k\in \mathbb{R}^{D_\mathbf{x}}$ denotes the $D_\mathbf{x}$-dimensional state, $\mathbf{y}_k\in \mathbb{R}^{D_\mathbf{y}}$ denotes the $D_\mathbf{y}$-dimensional observation (also called measurement), and $\mathbf{v}_k\in \mathbb{R}^{D_\mathbf{y}}$ denotes the observation noise. 

In contrast, there are different ways to model the target motion, which rests on the root of different estimation approaches. First, one may infer the state directly from the observation via maximum-likelihood estimation (MLE) or direct observation-to-state projection \cite{Li16O2, Li17clustering,Li17MSDC}, without relying on any statistical assumptions on the state process. This class of ``\textit{data-driven}" solutions will yield good results when the sensor data are highly informative (namely, the noise is very small), and are gaining favor when little is known about the motion model of the target, or when it is simply not interested/worthwhile or too hard to approximate one.  
Benefit to do so can particularly be seen in the context of visual tracking \cite{Smeulders14, Luo15} and chaotic time series \cite{Pisarenko04, Perretti13, Judd09failure}, where the target (e.g., pedestrian) motion is hard to be correctly/precisely modeled. However, in general, when there is any model information about the target dynamics, it should be carefully evaluated and utilized. This forms the majority of the existing efforts for estimation and the way to do so distinguishes our approach from the others. 

The prevailing, considered the standard, ``\textit{model-driven}" estimation solution is to apply a hidden Markov model (HMM) to link the target state over time, for online recursive computing. The best known methodology in this category is the sequential Bayesian inference, for which a filter consisting of \textit{prediction} and \textit{correction} steps is applied iteratively \cite{Li17AGC}. 
The HMM can be written in either discrete-time (mainly for convenience) or continuous-time (which is the nature of reality), as given by difference equation \eqref{eq:2a} and differential equation \eqref{eq:2b}, respectively, 
\begin{subequations} \label{eq:2}
\begin{equation} \label{eq:2a}
\mathbf{x}_k = g_k(\mathbf{x}_{k-1},\mathbf{u}_k),
\end{equation}
\begin{equation} \label{eq:2b}
\frac{d\mathbf{x}_t}{dt} = g_t(\mathbf{x}_{t},\mathbf{u}_t),
\end{equation}
\end{subequations}
where $t\in \mathbb{R}^+$ indicates continuous time, $\mathbf{u}_k \in \mathbb{R}^{D_\mathbf{x}}$ and $\mathbf{u}_t \in \mathbb{R}^{D_\mathbf{x}}$ denote the discrete time and continuous time state process noises, respectively. 

\subsection{Challenges to HMM}

\begin{itemize}
\item \textbf{Challenge 1} \textit{Difficulties involved in statistical modeling and full Bayes posterior computing, leading to inevitable model error and approximate computation, respectively}. 
\end{itemize}

All target models with parameters are no more than statistical simplification to the truth and inevitably suffer from approximation errors and disturbances. Challenges involved in system modeling/identification have been noted in several aspects, e.g., the model must meet practical constraints \cite{Li16, Duan16, Kurz16, Li17AGC
} and match the sensor revisit rate \cite{Li17AGC} while noises need to be properly identified \cite{Li15bias, Duník17,Ristic17}. In particular, the noise $\mathbf{u}_k/\mathbf{u}_t$ represents the uncertainty of the state process model, which has to be modeled with respect to the occasionally irregular revisit rate of the sensor (including missed detection, delayed or out of sequence measurements)
. Instead, there are also a few works using a deterministic Markov transition model \cite{Morrison2012tracking, Nadjiasngar13,Judd00, Judd08, Judd09failure, Judd09forecasting, Judd15, Smith10} which does not define the state process noise/uncertainty item; 
see further discussion given in Section \ref{literature}. In fact, in the majority of practical setups, the ground truth is deterministic conditioned on which, the Bayesian Cram\'{e}r-Rao lower bounds (CRLB) do not provide a lower bound for the conditional mean square error (MSE) matrix \cite{Fritsche16} and so correspondingly, no Bayesian filters can yield Minimum MSE estimate. 

One of the main reasons for the popularity of HMMs is the friendly assumption that states are conditionally-independent given the previous state. This allows easy forward-backward recursive inference, namely prediction-smoothing, but also severely limits the temporal dependencies that can be modeled which invites many alternatives \cite{Layton06,Frigola14,Bielecki17}. However, recursive estimators of the prior-updating format are vulnerable to the prior bias. Once an estimate bias is made, whether due to erroneous modeling, disturbances or over approximation, it will propagate in recursions and can hardly be removed \cite{Li16O2} unless a sliding time window or fading factor is used to adjust or re-initialize the estimator. A biased prior will likely not benefit the filter, especially when the observation is of high quality but instead, the filter may perform worse than the observation-only (O2) inference \cite{Li16O2, Li17clustering}. This fact however has been overlooked. While ``it is hard to overstate the importance of the role of a good model" \cite{Li03} for any model-driven tracker, model validation \cite{Djuric10, Thygesen17} has few been investigated.  


Even if the statistical models, uncertainties and constraints regarding the target (quantity and dynamics), sensor profiles (e.g., missed detection, clutter) and the scenario background can all be well approximated, the full Bayesian computation (or even the likelihood alone \cite{Tran15, Hartig11}) that involves integration over an infinite state space could be prohibitively expensive, forcing the need for further approximation
. This crucial challenge to real time computation and sensitivity to model-error will be escalated with the increasing revisit rate and joint use of advanced sensors \cite{Li17clustering,Li17MSDC}. 

\begin{itemize}
\item \textbf{Challenge 2} \textit{Strictness of the Markov-Bayes iteration that works on exact model assumptions and uses only limited statistical information while omitting others such as linguistic/fuzzy information}. 
\end{itemize}

Most model-driven estimators may work promisingly, e.g., minimizing the MSE, provided that the model assumptions come proper. However, many unexpected issues can occur in reality such as false, missing, morbid, biased, disordered data that are intractable for modeling and entail additional robust or adaptive processing schemes. This has formed the majority of extensions of the model-driven framework, e.g., a huge number of works for noise covariance estimation \cite{Duník17}, but also formed new challenges to the real time implementation due to escalated computational requirement. More importantly, it is unclear how to optimally use some important but fuzzy information such as a linguistic context that 
\textit{the target moves close to a straight line}, which might not be easily defined as constraints \cite{Duan16, Kurz16, Ko07, Julier07, Ravela07} and \cite[Chapter 6]{Sayed08}. This class of information is very common and useful to targets like aircraft, satellites, large cruise ships, and trains, which are supposed to move on pre-defined runways.

To alleviate these problems while gaining higher algorithm flexibility, we now substitute the stepwise Markov transition model \eqref{eq:2} with a deterministic FoT for describing the target motion. The resulting trajectory function that best fits the online sensor data series allows joint filtering, smoothing and forecasting while requiring fewer assumptions on the target motion and the background.

\subsection{Main contribution of our work}

At the core of our approach is a continuous-time trajectory function which is used to replace HMM for describing the target dynamics, i.e.,

\begin{equation} \label{eq:3}
\mathbf{x}_t=f(t).
\end{equation}

Combining the observation function \eqref{eq:1} and the trajectory FoT \eqref{eq:3} leads to an optimization problem for minimizing the sensor data fitting error. This can be written as
\begin{equation} \label{eq:4}
\underset{F(t)}{\text{argmin}} \sum_{t=k_1}^{k_2}\parallel \mathbf{y}_t-h_t(F(t),\bar{\mathbf{v}}_t)\parallel ,
\end{equation}
where $\parallel \mathbf{a}-\mathbf{b}\parallel$ is a measure of the distance between $\mathbf{a} \in \mathbb{R}^{D_\mathbf{y}}$ and $\mathbf{b} \in \mathbb{R}^{D_\mathbf{y}}$ such as the square error as in the least squares estimation, $F(t)$ is the FoT to be estimated based on the data provided in the underlying time-window $[k_1, k_2]$ and $\bar{\mathbf{v}}_t$ is an average to compensate for the observation error (if anything is known) and can be specified as the noise mean $\mathrm{E}[\mathbf{v}_t]$ if known or otherwise as zero by assuming the sensor unbiased. Since the estimate $F(t)$ needs to be updated at each time $k$ when a new observation is received, we denote the FoT obtained at time $k$ as $F_k (t)$. As the default, 
$k_2=k$, ensuring that the newest observation data are used in the fitting. 

To incorporate useful model information such as a linguistic description that ``\textit{the target is free falling}" or ``\textit{the target passes by a station}", the FoT can be more definitely specified as $F(t;C_k)$ of an engineering-friendly format such as a polynomial, with certain parameters $C_k$ to be estimated which reflects the a priori model information and fully determine the FoT at time $k$. To this end, the formula \eqref{eq:4} may be extended to a \textit{constrained} version, such as
\begin{align}
\begin{split} \label{eq:5}
\underset{F(t;C_k)}{\text{argmin}} & \sum_{t=k_1}^{k_2}\parallel \mathbf{y}_t-h_t(F(t;C_k),\bar{\mathbf{v}}_t)\parallel, \\
\text{s.t.} \hspace{0.5cm} & F(t;C_k) \in {\mathfrak{F}},
\end{split}
\end{align}
where $\mathfrak{F}$ is a finite set of specific functions, such as \textit{a set of polynomials of no more than 3-order}. 

Another common strategy to integrate the model information into the optimization formula is to additionally define a penalty factor $\Omega (C_k)$  on the model fitting error as a measure of the disagreement of the fitting function to the model constraint a priori. For instance, one can define $\Omega (C_k):=\parallel F(t_0;C_k)-\mathbf{x}_0\parallel$ to measure the mismatch between the fitting trajectory and the known state $\mathbf{x}_0$ that the target passes by/close-to at time $t_0$. Then, the formula \eqref{eq:4} is extended to 
\begin{equation} \label{eq:6}
\underset{F(t;C_k)}{\text{argmin}} \sum_{t=k_1}^{k_2}\parallel \mathbf{y}_t-h_t(F(t;C_k),\bar{\mathbf{v}}_t)\parallel + \lambda \Omega (C_k) ,  
\end{equation}
where $\lambda>0$ controls the trade-off between the data fitting error and the model fitting error. 

In this paper, we focus on a class of realistic targets of significance including the passenger aircraft or ship that moves in well-designed smooth routines (aside of which the tracker is given no other statistical information) and ballistic targets that are subject to the (nearly) constant velocity, acceleration or turn-rate. The challenge arises as that no quantitative statistical information is available about the target dynamics. 
Therefore, we will not explicitly define a quantized penalty function $\Omega(C_k)$ to account for the model fitting error at present, but instead directly assume the trajectory being a FoT of specified format, e.g., a polynomial, to reflect the a priori fuzzy information of ``\textit{smoothness}". 

Our earlier work \cite{Li16fitting} presented the idea of constructing the target trajectory FoT by fitting the discrete-time estimates yielded by an off-the-shelf estimator such as a Markov-Bayes filter or an O2/C4F estimator. In the latter the sensor data have to be converted to the state space which requires the observation model to be injective or multiple sensors available. In this paper, we ease this requirement and carry out fitting directly on the time-series observations rather than their conversion to the state space. The present approach is applicable to any observation model and does not make Markov-transition assumption. This is an essential difference of our work to the state of the art and our previous work. 

\section{Spatio-Temporal Trajectory Model} \label{fot}

In this paper, we limit our discussion to the state variables that can be observed directly, e.g., the target position, rather than all variables of interest such as position, velocity and acceleration. To clarify this, an important definition is made on the concept of ``\textit{directly-observed state}". 

\textbf{Definition 1} \textit{Directly-observed state} The ``directly-observed state" is referred to the variables of the state that are deterministic function of the observation. For example, for the range and bearing observation, the directly-observed state is the target position while for the Doppler observation, the directly-observed state is the radial velocity. 

We note that some unobserved variables of interest may be inferred from the directly-observed variables based on the laws of physics, e.g., the differentiation of position and velocity over time is velocity and acceleration, respectively. 

\subsection{General framework}
Instead of \eqref{eq:2}, we propose using the trajectory FoT \eqref{eq:3} to model the target motion. Our goal is to find a engineering-friendly $F(t)$ for approximating $f(t)$ which best fits the sensor data as in \eqref{eq:4} and then use it to estimate the state for a desired time. 
To avoid the computationally intractable hyper-surface fitting, we perform fitting with respect to each dimension of the ``\textit{directly-observed state}", by assuming conditional independence among the dimensions. This is computationally efficient since neither integration nor differentiation in one dimension will affect the others in the orthogonal coordinates.  

Any continuous trajectory can be approximated by a polynomial of a certain degree to an arbitrary accuracy \cite{Bar-Shalom01, Li03}. This accuracy can be easily analyzed by the Taylor series expansion as in Appendix A. Based on this, linear parameter dependence can be assumed, 
i.e., 
\begin{equation} \label{eq:7}
F(t)=c_1\phi_1(t)+c_2\phi_2(t)+\cdots+c_m\phi_m(t) ,
\end{equation}
where $\{\phi_i(t)\}$ are a priori selected sets of functions, for example, monomials $\{t^{i-1}\}$ or trigonometric functions $\{\text{sin}it\}$, and $C:=\{c_i\}, i=1,2,…,m$ are parameters to be determined. We call $m$ the order of the fitting function, which controls the number of free parameters in the model and thereby governs the model complexity. 

Hereafter, we denote the parameterized FoT $F_k (t)$ in a specified form $F(t;C_k)$, where $C_k$ denotes the parameter set at the time $k$. 
It is crucial to determine the order $m$ of the polynomial properly. 
For the typical CV and CA models, we have the following constraint, respectively,
\begin{equation} \label{eq:8}
\frac{\partial f_\text{CV}(t)} {\partial t} = \text{Constant}, \frac{\partial^2 f_\text{CA}(t)} {\partial^2 t} = \text{Constant} ,
\end{equation}
which indicates that the suitable fitting function order for the CV and CA motions are $m=2$ and $m=3$, respectively.

In our approach, we advocate sliding time-window fitting (to be detailed in Section \ref{Time-window}), by which any trajectory function can be divided into a number of consecutive time intervals, each of which corresponds to one function of relatively lower order. This allows us trade-off fitting fidelity to the real trajectory FoT with computational 
complexity. 
For example, a two dimensional polynomial with $m=2$ given as follows applies to most \textit{smooth} trajectories in the planar space  
\begin{equation} \label{eq:9}
F(t,C_k): \left[ \begin{array}{c}
x(t) \\
y(t) \\
\end{array} \right]=\left[ \begin{array}{ccc}
a_{k,1} & a_{k,2} & a_{k,3} \\
b_{k,1} & b_{k,2} & b_{k,3} \\
\end{array} \right] \left[ \begin{array}{c}
1 \\
t \\
t^2 \\
\end{array} \right] ,
\end{equation}
where 
$C_k:=\{a_{k,i},b_{k,i}\}_{i=1,2,3}$. 

While the above formulation is suitable to fit straight lines or smooth curves, the trigonometric function is particularly useful for circular trajectories, e.g., an elliptic trajectory in the planar Cartesian space can be represented as 
\begin{equation} \label{eq:10}
a'_{k,2} \big(x(t)-a'_{k,1}\big)^2+b'_{k,2} \big(y(t)-b'_{k,1}\big)^2= 1,
\end{equation}
for which $C_k:=\{a'_{k,i},b'_{k,i}\}, i=1,2$.

The elliptic curve can also be given by a parametric form with the $x$ and $y$ coordinates having different scalings 
\begin{equation} \label{eq:11}
F(t;C_k ):\left\{
\begin{array}{ll} 
x(t)=a_{k,1} + a_{k,2} \text{sin}⁡ (t)\\
y(t)=b_{k,1} + b_{k,2} \text{cos⁡} (t)
\end{array} \right. ,
\end{equation}
and correspondingly $C_k:=\{a_{k,i},b_{k,i}\}, i=1,2$.

\subsection{FoT Parameter Estimation and Optimization}
To get the desired FoT $F(t;C_k)$, the parameters $C_k$ can be determined by minimizing the fitting residual $\parallel F (t;C_k)-f_k (t)\parallel$ in the underlying time window $[k_1,k_2]$. However, these residuals are not explicitly available since the true trajectory FoT $f_k(t)$ is unknown and is exactly what we want to estimate. As such, we turn to selecting the function that best fits the sensor data series as in \eqref{eq:4} in which the fitting residual is defined by the discrepancy between the original sensor data and the pseudo observation made on the FoT of the corresponding time, namely, 
\begin{equation} \label{eq:12}
R_t(C_k):=\parallel \mathbf{y}_t-h_t(F(t;C_k),\bar{\mathbf{v}}_t)\parallel .
\end{equation}

Typically, the distance $\parallel \cdot\parallel$ can be given as the un-weighted $\ell_2$-norm of the error (square error), namely $R_t(C_k)|_{\ell_2}:=\parallel \mathbf{y}_t-h_t(F(t;C_k),\bar{\mathbf{v}}_t)\parallel^2$. This is referred to as the least squares (LS) fitting for which the Gauss-Newton method \cite{Morrison2012tracking} is popular. 
The resulting $\hat{F}(t;C_k)$ is known as the ordinary LS fitting to the given data which implicitly assumes that the sensor data received at different time instants provide equally accurate information. 

Further on, one can assign weight $w_t$  to each time instant to account for time-varying data uncertainty, e.g., being inversely proportional to the covariance of $\mathbf{y}_t$, leading to
\begin{equation} \label{eq:13}
R_t(C_k)|_{\ell_2}:=\tilde{w}_t\parallel \mathbf{y}_t-h_t(F(t;C_k),\bar{\mathbf{v}}_t)\parallel^2 ,
\end{equation}
where $\tilde{w}_t=w_t\big(\sum_{t=k_1}^{k_2} w_t\big)^{-1}$ is the normalized weight regarding all data in the time window. Moreover, a fading factor can also be considered in the weight design, such as $w_t:=\lambda^{k-t}$ where $0<\lambda<1$, in order to emphasize the newest data by assigning lower weights to history data.

We denote by $ \Phi_k(C_k):=\sum_{t=k_1}^{k_2}R_t(C_k)$ the sum of the residuals in the time window $[k_1,k_2]$. Then, the fitting problem is reduced to parameter estimation of $C_k$, namely,
\begin{equation}\label{eq:14}
\underset{C_k}{\text{argmin}}\Phi_k(C_k) ,
\end{equation}
where the sensor data $\mathbf{y}_t$  may arrive at irregular time intervals and suffer from missed detection and outlier/false alarms. 

\subsection{Analytic Solution and Numerical Approximation}
In general, the (necessary) condition for $\Phi_k(C_k)$  to be a minimum is that the following $m$ gradient equations are zero, namely, 
\begin{equation} \label{eq:15}
\frac{\partial \Phi_k(C_k)} {\partial c_i} = 0, \forall i=1,...,m.
\end{equation}

In a linear system (and for continuously increasing time window, namely $k_1=0$), it can be exactly solved given sufficient sensor data, e.g., by a recursive LS algorithm \cite[pp.138] {Gustafsson00}, \cite[Chapter 10]{haykin14}, \cite[Chapter 30]{Sayed08} as briefly shown in Appendix B. Particularly, when equality constraints exist, they can be easily incorporated into the minimization function by methods such as Lagrange multipliers \cite{Lagrange}. 

However, in a nonlinear system, the derivatives are functions of both the independent variable and the parameters, which make these gradient equations do not have a closed solution. Instead, we have to resort to numerical approximation methods such as the trust-region-reflective (TRR) algorithm \cite{Sorensen97, Branch99} and the Levenberg-Marquardt algorithm (LMA) \cite{Kanzow04}. In particular, constraints on parameters, e.g. bounded parameters, can be easily integrated in TRR \cite{Branch99}, as in \eqref{eq:5}. These algorithms have been implemented in popular software and compute efficiently, offering great convenience for engineering use. However, we must note that almost all fitting algorithms including TRR and LMA work from an initial guess of the parameters for iterative searching and do not guarantee finding the global minimum. That is, the parameters are obtained by successive approximation till the residuals $\Phi_k$ do not decrease significantly in iterations or become lower than a threshold. 

To speed up the iteration, we set the parameters $C_{k-1}$ yielded at time $k-1$ as the initial parameters for estimating $C_k$ at time $k$. This is feasible because the trajectory functions yielded by the data in time window $[k_1,k_2]$ and that in time window $[k_1+1,k_2+1]$ will be similar due to the common data in $[k_1+1,k_2]$. In this setting, the result is a recursive algorithm for which the recursion can be described as 
\begin{equation} \label{eq:16}
C_k = \Psi_k(C_{k-1}) .
\end{equation}

It is worth noting that the parameter transition in \eqref{eq:16} due to data updating in the sliding fixed-length time window is not a parameter convergence process as in the recursive LS estimation due to new data adding to the continuous increasing time window therein. It is also not necessarily a Markov-jump process as the parameters at time $k$ depends not only on that at time $k-1$ but also that of earlier times, if the fitting interval is longer than unity. 

To ameliorate the computational complexity due to nonlinear fitting, an alternative method is to project the sensor data into the state space as is done by the O2 approach \cite{Li16O2, Li17clustering, Li17MSDC}. Then, the problem reduces to performing linear fitting on the intermediate O2 estimates $\hat{x}^\text{O2}_{k_1:k_2} $ for which the fitting residual is given by $R_t(C_k):=\parallel \hat{x}^\text{O2}_t-F(t;C_k)\parallel $. This however requires the observation function being injective or multiple sensors being used which does not apply to the sensor setup such as a bearing-only sensor. 

\section{Trajectory FoT Based STF} \label{stf}
Given an FoT estimate $F(t;C_k)$, the state at any time $t$ (that does not have to be an integer) in the effective fitting time window (EFTW) $[K_1,K_2]$ can be estimated as follows
\begin{equation} \label{eq:17}
\hat{\mathbf{x}}_t=F(t;C_k), \forall t\in [K_1,K_2],
\end{equation}
where EFTW $[K_1,K_2]$ at least covers the sampling time window $[k_1,k_2]$, namely $K_1 \leq k_1, k_2\leq K_2$. More specifically, the inference is referred to as extrapolation when $K_1\leq t <k_1$ or $k_2 \leq t \leq K_2$ and as interpolation when $k_1\leq t\leq  k_2$. Different choices of time $t\in [K_1,K_2]$ with regard to the right bound of the time window $k_2$, are immediately apparent and correspond to different fitting terminologies as follows: 

\begin{itemize}
\item \textbf{Delayed fitting}: $t<k_2$, the state to infer is for an earlier time. Particularly, we notice that fitting at the middle of the time window, namely $t=(k_1+k_2 )/2$, is comparably more accurate. This is also referred to as \textit{fixed-lag smoothing} as the estimation bears a fixed time-delay of $k_2-t$. 
\item \textbf{Online fitting}: $t=k_2$, the state to infer/filter is exactly for the time when the latest sensor data arrive.
\item \textbf{Forecasting}: $t>k_2$, the state to infer is for the future time. Particularly, denoting $n:=(t-k_2)$, it is called $n$-step forecasting. 
\end{itemize}

In addition, any estimates given above can be further fitted over a time window, forward and backward in time series, as many times as desired, to repeatedly revise the fitting function for more accurate trajectory estimation. This type of batch/off-line fitting is referred to as \textbf{Smoothed Fitting} hereafter. This is preferable for off-line data analysis but caution should be exercised to avoid over-fit. We take once-forward and once-backward delayed fitting as the default smoothed fitting, which resembles the conventional \textit{fixed-interval smoothing}. It can be written as
\begin{equation} \label{eq:18}
\underset{C_k}{\text{argmin}} \sum_{t=k_1}^{k_2}\parallel \hat{x}^\text{DF}_t-F(t;C_k)\parallel ,
\end{equation}
where $\{\hat{x}^\text{DF}_t\}, t=k_2,k_2-1,\cdots,k_1$ are the estimates yielded by the \textit{Delayed Fitting} and the fitting time-window $[k_1,k_2]$ moves in reverse order by time (namely, backward). 

Finally, we emphasize two important points about fitting, which are particularly beneficial for our approach.

\subsection{Piecewise/Sliding Time-window Fitting} \label{Time-window}
Numerical fitting over a long data series suffers from instability, especially when the trajectory is subject to different models at different time periods. In such a situation, piecewise fitting, also referred to as spline fitting or segmented fitting, is a useful alternative.  
The advantage of piecewise fitting is that at each time instant, the complexity of the fitting function can be controlled (of lower order), which will not be affected by the data outside the time window. At the core of piecewise fitting is to detect the model change from the sensor data in time series, where the change point is the desired boundary between segments. This is formally known as \textit{change-point detection} in general\cite{Gustafsson00} or \textit{maneuver detection} in the context of target tracking \cite{Li03, Ru09, 
Li17Maneuvering}. There are a large body of algorithms and softwares; see e.g., \cite{Gustafsson00, Ross15}. 

However, we want to indicate that most change detection mechanisms, including our own previous attempt \cite{Li16fitting}, are problem-dependent and suffer from detection delay. In this paper, sliding time-window fitting (which is a special type of piecewise fitting) is advocated. With a sliding time-window fitting, the time-window $[k_1,k_2]$ is supposed to move forward with time $k$. 
The length of the time window can be adapted to accommodate high varying target dynamics, in accordance to the order of the fitting function and the feasible computing time that has to be smaller than the sampling interval. For the target we consider here such as passenger aircraft/ships and satellites, even a maneuver occurs, the target trajectory may remain smooth. 
In this case, maneuver detection becomes unnecessary as the fitting can be carried out the same when the target maneuvers smoothly, i.e., the trajectory remains smooth as 
demonstrated in our simulations in Sections \ref{sec:linear} and \ref{sec:nonlinear}. We have particularly addressed this issue in \cite{Li17Maneuvering}. 
It is one of the advantages yielded by formulating the target motion as a FoT rather than by a Markov transition model. 

\subsection{Missed detection and irregular sensor data}
Fundamentally different from the Markov-based estimator that needs to assume independence among time series states, time-window fitting eases such assumptions and does not require the data to be uniformly observed over time or to be chronological. Because of this, neither missed detection/delayed data, nor irregular sensor revisit frequency will inhibit our approach so much as it does to a Markov-Bayes estimator. 

In fact, both missed detection and delayed observation can be viewed as a special case of the problem of sensor data arriving at irregular time intervals, which do not constitute any challenge to fitting as long as the sensor data and its corresponding time are correctly matched. This greatly adds to the flexibility and reliability of our approach. Next, we will review work on trajectory estimation, some of which are based on fitting, and exhibit advantages for coping with the missed detection and irregular sensor data. 

\section{Related Work}  \label{relatedwork}

Target trajectory estimation and analysis
do not only allow recording the history of past locations 
and predicting the future 
but also provide a means to specify behavioral patterns or relationships between locations observed in time series and to guide target detection in future frames \cite{Leibe07}, to name a few. 
Most existing works however are based on either deterministic or stochastic HMM assumption of the target motion and need statistical property of the observation, which forms the key difference to our approach. In addition, 
no explicit attempts explicitly unify the tasks of smoothing, filtering, tracking and forecasting, fully based on data fitting/learning.

\subsection{Discrete-time trajectory estimation} \label{literature}

Instead of estimating point-states at each time instant when a new observation is received, there are some studies that recursively estimate the discrete time-series state set \cite{Guerriero10, Smith10, García-Fernández16} based on a sequence of observations. 
Compared to the recursive point-state estimation, this, as generally termed data assimilation when formulated as an optimization problem, requires much higher computation. For linear systems, discrete-time trajectory estimation has direct connection to Gauss' LS estimate \cite{Plackett50} and Kolmogorov-Wiener's interpolation and extrapolation of a sequence \cite{Singpurwalla17}. 
Data assimilation refers to finding trajectories of a prescribed dynamical model such that the output of the model follows a given time series of observations \cite{Dimet86, Talagrand87, Bröcker12, Wang13, Rosenthal17}. The key point is to search for the maximum of the posterior density function by assuming certain (e.g., Gaussian) observation noise, model initial conditions and model errors, and iteratively minimizing a cost function which is the negative of the logarithm of the posterior density function.
Of high relevance, an expectation maximization (EM) approach is proposed \cite{Shumway82} in conjunction with conventional Kalman smoothers for smoothing and forecasting, yielding a recursive procedure for estimating the parameters by MLE, which can deal with missing observations. 

Differently to the stochastic modeling of the state process, Judd etc. presented a series of non-sequential/optimization-based estimation and forecasting works, particularly in the area of chaotic systems, e.g., \cite{Judd00, Judd08, Judd09forecasting, Judd15, Smith10}, which 
remove the use of the state transition noise. Actually, similar deterministic Markov models have been applied in noise reduction methods \cite{Kostelich93}, moving horizon estimator \cite{Michalska95} and Gauss-Newton filter \cite{Morrison2012tracking, Nadjiasngar13}. Interestingly, Judd's shadowing filter yields more reliable and even more accurate performance than the Bayesian filters - \textit{however, a fairer comparison should be made between shadowing filters with Bayesian smoothers, using the same amount of observation data} - in the case when the nonlinearity is significant, but the noise is largely observational \cite{Judd09failure}, or when the objects do not typically display any significant random motions at the length and the time scales of interest \cite{Judd15}. The Gauss-Newton filter that models the state transition by a deterministic differential equation, namely \eqref{eq:2b} without noise $\mathbf{u}_t$, is Cram\'{e}r-Rao consistent (providing minimum variance)\cite{Morrison2012tracking}. Despite their Markov assumptions, these approaches, similar to our fitting approach, are based on optimization formulation, which is advantageous in handling constraints (as shown in \eqref{eq:33} in Section \ref{simulation}.C) and is less sensitive to process disturbances, missing data and observation singularities than a recursive Bayesian filter.
 
\subsection{Continuous-time trajectory estimation}

More relevantly to our approach, efforts have been devoted to continuous time trajectory estimation via data fitting in different disciplines
. De facto, signal processing stems from the interpolation and extrapolation of a sequence of observations
 \cite{Singpurwalla17}. Data fitting is a self-contained mathematical problem and a prosperous research theme by its own, which has proven to be a powerful and universal method for pattern learning and time series data prediction, especially when adequate analytical solutions may not exist. Moreover, the recursive LS algorithm reformulated in state-space form was recognized a special case of the Kalman filter (KF)\cite{Sorenson70,Sayed94}.

However, most existing works work in batch manners based on either MLE \cite{Anderson-Sprecher96} or Bayesian inference \cite{Hadzagic11, Dimatteo01} or as an extra scheme to a recursive filtering algorithm \cite{El-Hawary95,Wang10,Liu14}. In \cite{Anderson-Sprecher96}, directional bearing data from one or multiple sensors are investigated, where Cardinal splines (i.e., splines with equally spaced knots) of different dimensions are fit to the data in the MLE manner; this is one of the earliest and few attempts that assume a spatio-temporal trajectory for tracking. In \cite{Hadzagic11}, the trajectory is approximated by a cubic spline with an unknown number of knots in 2D state plane, and the function estimate is determined from positional measurements which are assumed to be received in batches at irregular time intervals. For the data drawn from an exponential family, the spline knot con-figurations (number and locations) are changed by reversible-jump Markov chain Monte Carlo \cite{Dimatteo01}. Much more complicated, artificial neural networks were considered as a parametric non-linear model in \cite{Hamed13}, which is unaffordable in computation for online estimation. 

Continuous time trajectory estimation has also received attention in the context of mobile robot simultaneous localization and mapping (SLAM) \cite{Bibby10, Lovegrove13} and visual tracking \cite{Delong10, Milan16}. In the former, the robot motion is usually under the user's control (called proprioceptive sensor data) and the continuous-time trajectory representation makes it easy to deal with asynchronous measurements and constraints. In the latter, starting and/or ending points may be specified for the trajectory. 
The tracking problem is treated as the discrete-continuous optimization with label costs\cite{Milan16}, where the key is generating all the trajectory hypotheses having a reasonable low label cost based on a variety of DA rules, for which the design of the label cost takes the critical issues such as the targets' dynamics, occlusions and collisions into account. However, only linear fitting is involved. 

Parametric curve fitting methods have the difficulty to define knots. Comparably, Gaussian process (GP) provides a non-parametric tool for learning regression functions from data, having the advantage of providing uncertainty estimates \cite{Anderson15} under linear state process function. Furthermore, regression based on a support vector regression model and a GP model respectively was advocated to predict ballistic coefficients of high-speed vehicles and also the long-term future state \cite{Song16}. Relevantly, the Gaussian smoothing \cite{Särkkä13} allows for inferring the state at any time of interest using the interpolation scheme that is inherent to GP regression. Multi-step ahead prediction in time series analysis is treated using the non-parametric GP model \cite{Girard03} in the manner of making repeated one-step ahead predictions. There methods are all based on data training and again, stepwise state transition.

In summary, our FoT fitting approach differs from the above various data fitting approaches (not only for target tracking) in four major aspects:
\begin{itemize}
\item The assumed trajectory function is purely a function of continuous time (namely spatio-temporal, ``$x=f(t)$"), rather than a spatial function defined in the state space in a manner like ``$y=f(x)$";
\item We perform continuous time fitting in each state-dimension independently;
\item We use a sliding time window rather than pre-defined, ad-hoc knots for fitting flexibility; and
\item Our approach accommodates complete absence of statistical information about the target and the sensor.
\end{itemize}

Moreover, we reiterate two critical, original strategies for efficiently real time implementation of our approach
\begin{itemize}
\item As indicated by \eqref{eq:8}, usual description of the target motion can be utilized for fitting function design; and
\item Thanks to the use of \eqref{eq:16} in nonlinear fitting initialization, our approach can be carried out online.
\end{itemize}


\section{Simulation}  \label{simulation}

Although the proposed FoT formulation of the target motion is fundamentally different from HMM, 
it is interesting to compare the FoT-based STF approach with the state-of-the-art Markov-Bayes solutions. For this, this section will study a variety of representative scenarios. In all cases, our fitting approach does not need to make any statistical assumption on the target dynamics, background or sensor noise while ideal statistical information is provided to the Markov-Bayes filters, smoothers and forecasters except otherwise stated for their most favorable performance. 

For the sake of generality and reproduction of the results, the first two simulations are taken from an excellent Matlab toolbox due to Hartikainen, Solin, and S\"{a}rkk\"{a} \cite{Hartikainen13}: one uses linear and non-deterministic target dynamics and linear observation model, while the other utilizes deterministic target dynamics and nonlinear observation model. This toolbox features a large body of popular filters and smooths for discrete-time state space models, including the KF, extended KF (EKF) and unscented KF (UKF) and their corresponding smoothers implemented on the basis of the rauch-tung-striebel (RTS) algorithm. In addition, the interacting multiple model (IMM) approach, as well as its non-linear extensions based on the mentioned filters and RTS smoothers, has also been simulated. In contrast, the third simulation is described in a continuous-time system for tracking a non-maneuvering ballistic target in which a particle filter (PF) is compared. 

The Matlab codes used for the simulation are available at: {\small https://sites.google.com/site/tianchengli85/matlab-codes/fot4stf}

\subsection{Linear observation maneuvering target tracking} \label{sec:linear}

This simulation example is the same as that described in Section 4.1.4 of \cite{Hartikainen13}, where the motion of a maneuvering object switches between WPV (Wiener process velocity) with low process noise, such as a power spectral density $0.1$, and WPA (Wiener process acceleration) with high process noise, such as a power spectral density $1$. The system is simulated with 200 sampling steps (with the sensor revisit interval $\Delta= 0.1$s). The real target motion model was manually set to WPV during steps 1-50, 71-120 and 151-200 and to WPA during steps 51-70, and 121-150. This leads to four maneuvers. 

The initial state of the target is $\mathbf{x}_0  = [0, 0, 0, -1, 0, 0]^T$, which means that the object starts to move from the origin with velocity $-1$ along the y-axis. All filters and smoothers are correctly initialized with the true origin $\mathbf{x}_0$ and the covariance diag$([0.1,0.1,0.1,0.1,0.5,0.5]^T )$. In addition, for the best possible performance of the IMM approach, correct knowledge about the two models is assumed (except the maneuvering time). The prior model probabilities are set to $[0.9,0.1]^T$ and the model transition probability matrix for IMM is set to
\begin{equation} \label{eq:19}
Tr_\text{IMM}=\left[ \begin{array}{cc}
0.98 & 0.02 \\
0.02 & 0.98 \\
\end{array} \right] .
\end{equation}

Observation $\mathbf{y}_k$ is made on the target position $[p_{x,k},p_{y,k}]^T$ with Gaussian noise $\mathbf{v_k}$, $\mathbf{y}_k = [p_{x,k},p_{y,k}]^T +\mathbf{v}_k$ where
\begin{equation} \label{eq:20}
\mathrm{E}[\mathbf{v}_k]=\mathbf{0}, 
\mathrm{E}[\mathbf{v}_k\mathbf{v}_j^T]=\left[ \begin{array}{cc}
0.1 & 0 \\
0 & 0.1 \\
\end{array} \right]\delta_{kj},
\end{equation}
where $\delta_{kj}$ is the Kronecker-delta function which equal to one if $k=j$ and to zero otherwise. 

\begin{figure*}\label{fig:1}
\centering
\includegraphics[width=14.2 cm]{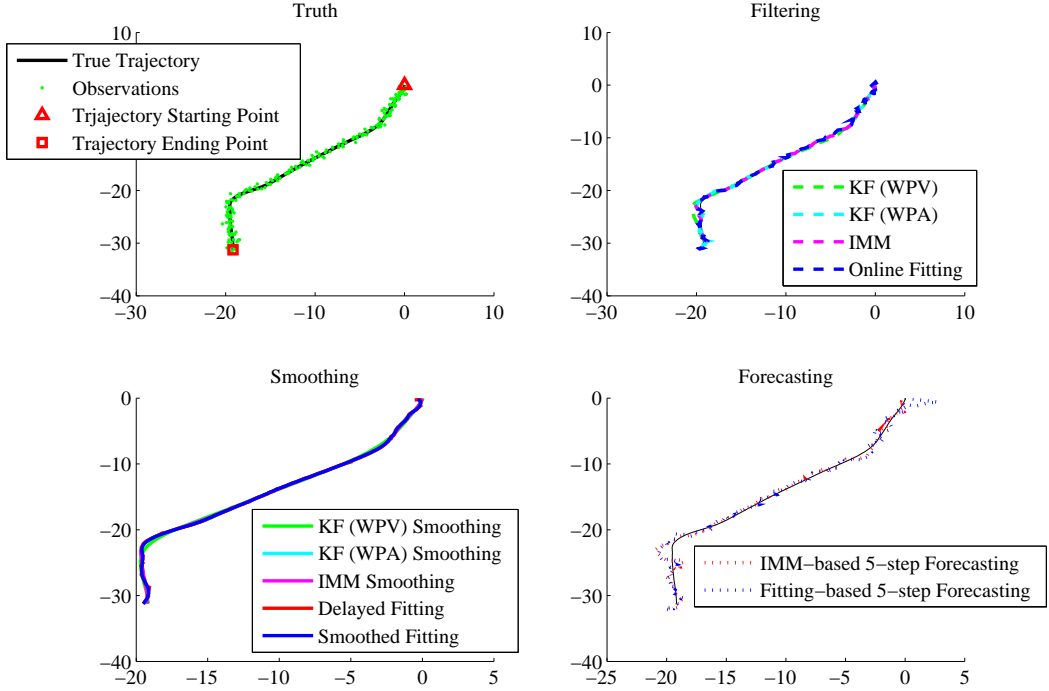}
\caption{Real trajectory and observations of one trial, and estimates given by different linear estimators (filters, smoothers and forecasters respectively)}
\end{figure*}
\begin{figure}\label{fig:2}
\centering
\includegraphics[width=8 cm]{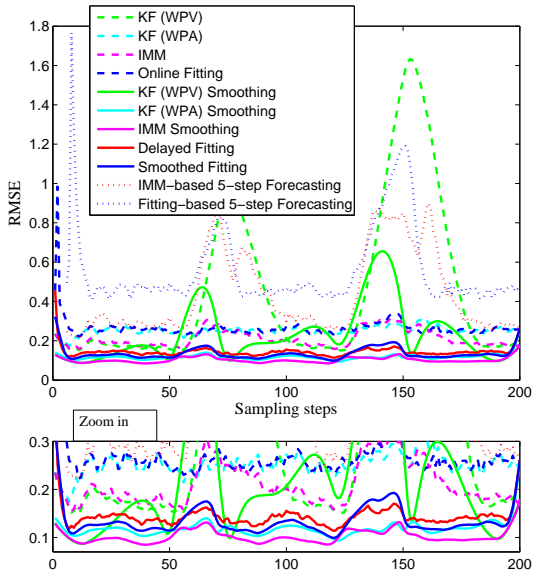}
\caption{RMSE of different linear estimators over sampling steps}
\end{figure}

For this unbiased linear measurement model, the unbiased O2 position estimates can be directly given by measurement $[p^\text{O2}_{x,k},p^\text{O2}_{y,k}]^T=\mathbf{y}_k$.

The proposed trajectory FoT fitting is carried out in the x-axis and y-axis individually in the LS manner, with a sliding time window of 10 sampling steps (except the starting stage when little data are available). The polynomial trajectory FoT of order $m=2$ is assumed as follows
\begin{equation} \label{eq:21}
\left\{
\begin{array}{ll} 
p_{x,t}=a_1+a_2t\\
p_{y,t}=b_1+b_2t 
\end{array} \right. ,
\end{equation}
and the optimization goal is given as
\begin{equation} \label{eq:22}
\left\{
\begin{array}{ll} 
\Phi_k(a_1,a_2) := \sum_{t=k_1}^{k_2} \big( p^\text{O2}_{x,t}-(a_1+a_2t)\big)^2\\
\Phi_k(b_1,b_2) := \sum_{t=k_1}^{k_2} \big( p^\text{O2}_{y,t}-(b_1+b_2t)\big)^2
\end{array} \right. ,
\end{equation}
where $k_2$ is the current time, $k_1=\text{max}⁡(1,k_2-10)$.

Given a time series of O2 estimates $[p^\text{O2}_{x,k},p^\text{O2}_{y,k}]^T$ for the time window $k\in [k_1,k_2]$, the trajectory FoT parameter can be exactly determined by \eqref{eq:15}, i.e.,

\begin{equation} \label{eq:23}
\left\{
\begin{array}{ll} 
\sum_{t=k_1}^{k_2} \big(a_1+a_2t-p^\text{O2}_{x,t}\big)=0\\
\sum_{t=k_1}^{k_2} \big(a_1t+a_2t^2-p^\text{O2}_{x,t}t\big)=0\\
\sum_{t=k_1}^{k_2} \big(b_1+b_2t-p^\text{O2}_{y,t}\big)=0\\
\sum_{t=k_1}^{k_2} \big(b_1t+b_2t^2-p^\text{O2}_{y,t}t\big)=0\\
\end{array} \right. .
\end{equation}

Once $a_1,a_2,b_1,b_2$ are obtained, the position of the target at time $t$ can be inferred as $[\hat{p}_{x,t},\hat{p}_{x,t}]^T=[a_1+a_2t,b_1+b_2t]^T$ straightforwardly. As addressed, four forms of fitting-inference can be implemented based on the same fitting function: delayed fitting ($t=k_2-5\Delta$ which estimates the state with $0.5$s delay), online fitting ($t=k_2$ which estimates the state using the latest 10 sensor data), forecasting ($t=k_2+5\Delta$ which estimates the state of the future, 0.5s in advance), and smoothed fitting (which is given by carrying out the delayed fitting forward in time series and then backward). Very different from IMM approach, our approach needs neither any multiple-model design for dealing with the target maneuver nor any a priori knowledge of the initial target state for estimator initialization. This, at the starting point, reveals the robustness advantage of fitting. 

The simulation is performed with 100 Monte Carlo runs, each run having a randomly generated trajectory originating from the same initial point, and a corresponding independently generated observation-series. The real trajectory and the estimates given by different filters, smoothers and forecasters in one run are given in Fig.1. The IMM-based 5-step forecasting is given by iteratively carrying out the prediction of the IMM approach 5 times without observation updating. 
As shown in Fig.1, all estimators correctly capture the trend of the trajectory. For more insights, their root MSEs (RMSEs) on the position estimation over time are given in Fig.2, where RMSE
is calculated over 100 Monte Carlo runs. The mean of all $\text{RMSE}_k$ over the 200 sampling steps and the average running time per run are given in Table \ref{tab:1}. 
\begin{itemize}
\item On the estimation accuracy, the online fitting outperforms the KF based on the MPV model but underperforms the IMM and the KF using the MPA model. The smoothed fitting improves over the delayed fitting, both outperforming the KS (Kalman smoother) using WPV but losing to the KF using MPA and the IMM smoother. For 5-step ahead forecasting, our fitting approach outperforms the IMM approach. 
\item On the computing speed, the online fitting is slower than the filters, while the delayed fitting is slightly slower than the KSs using only one model, but is faster than the IMM smoother. The smoothed fitting is the slowest overall. However, the forecasting fitting is both faster and more accurate than the IMM forecaster. We must note here that all fitting approaches share the same fitting function, and their respective computing times have taken into account the common part for obtaining that fitting function (which is the majority of the computation required overall). Therefore, their joint computing time, if STF are required jointly, is much less than the sum of their respective times shown here. 
\end{itemize}

Although the estimation accuracy is slightly inferior to some of the suboptimal filters and smoothers that are based on ideal models and parameters, our fitting approaches work under the harsh condition that (i) they need neither a priori information about the target motion models nor the sensor observation noise statistics, but (ii) they provide powerful continuous-time estimates (yet we only compare the estimation at discrete time instants), including better prediction than that of model-based estimators
. More importantly, (iii) the proposed fitting approach obviates the need to design multiple/adaptive models due to target maneuver. 
However, when these filters and smoothers are not initialized perfectly with the true state and even with the ideal error covariance, and/or if the multiple-model approach is not designed properly, their performance will undoubtedly degrade. In contrast, the fitting scheme based on minimum unrealistic assumptions will not suffer from these problems. 
These are just the advantage of modeling the target dynamics by a trajectory FoT rather than by a HMM. 

Next, we will investigate two nonlinear systems, in which the estimators may be provided with incorrect sensor noise information or poorly initialized. 

\begin{table}[h]
\caption{Average performance of different linear estimators}
\label{tab:1}
\begin{center}
\begin{tabular}{|c||c||c|}
\hline
\textbf{Estimators} & \textbf{Aver. RMSE} & \textbf{Compt. Time (s)}\\
\hline 
KF (using WPV) & 0.4498	&0.0252\\
\hline
KS (using WPV)	&0.2308	&0.0472\\
\hline
KF (using WPA)	&0.2520	&0.0251\\
\hline
KS (using WPA)	&0.1184	&0.0496\\
\hline
IMM	&0.2116	&0.2864\\
\hline
IMM smoother &0.1025	&1.0875\\
\hline
IMM	5-step forecaster&0.4373	&0.9896\\
\hline
Online Fitting	&0.2654	&0.7120\\
\hline
Delayed Fitting	&0.1442	&0.7434\\
\hline
Smoothed Fitting &0.1348 &1.4686\\
\hline
Fitting-based 5-step Forecasting &0.5586 &0.7210\\
\hline
\end{tabular}
\end{center}
\end{table}

\subsection{Nonlinear observation maneuvering target tracking} \label{sec:nonlinear}
This simulation is set the same as that given in in Section 4.2.2 of \cite{Hartikainen13}. To simulate the deterministic target motion (as shown in Fig.1), two Markov models using insignificant noises are assumed with sampling step size $\Delta=0.1$s. The first is given by a single linear WPV model with insignificant process noise (zero-mean and power spectral density 0.01), based on which the standard EKF, UKF and their corresponding RTS smoothers (EKS and UKS respectively) are realized. The other is given by a combination of this WPV model with a nonlinear CT model (no position and velocity noise but zero-mean turn rate noise with covariance 0.15). In the latter, multi-model design and nonlinear estimation approaches are required for which the EKF/EKS-IMM and UKF/UKS-IMM approaches are employed for filtering/smoothing. The IMM uses a model transition probability matrix as follows
\begin{equation} \label{eq:24}
Tr_\text{IMM}=\left[ \begin{array}{cc}
0.9 & 0.1 \\
0.1 & 0.9 \\
\end{array} \right] ,
\end{equation}
with the prior model probabilities given by $[0.9,0.1]^T$.

The measurement is made on the noisy bearing of the object, which is given by four sensors located at $[s_{x,1},s_{y,1}]^T=[-0.5,3.5]^T$, $[s_{x,2},s_{y,2}]^T=[-0.5,-3.5]^T$, $[s_{x,3},s_{y,3}]^T=[7,-3.5]^T$ and $[s_{x,4},s_{y,4} ]^T=[7,3.5]^T$, respectively. The noisy bearing observation of sensor $i=1,2,3,4$ is given as 
\begin{equation} \label{eq:25}
\theta_{k,i}=\text{arctan}\Big(\frac{p_{y,k}-s_{y,i}}{p_{x,k}-s_{x,i}}\Big) + v_{k,i} ,
\end{equation}
where $v_{k,i} \sim \mathcal{N}(0,\Sigma_v)$ and we will use $\Sigma_v =0.01$ and $\Sigma_v=0.0025$ separately. 

\begin{figure}\label{fig:3}
\centering
\includegraphics[width=8 cm]{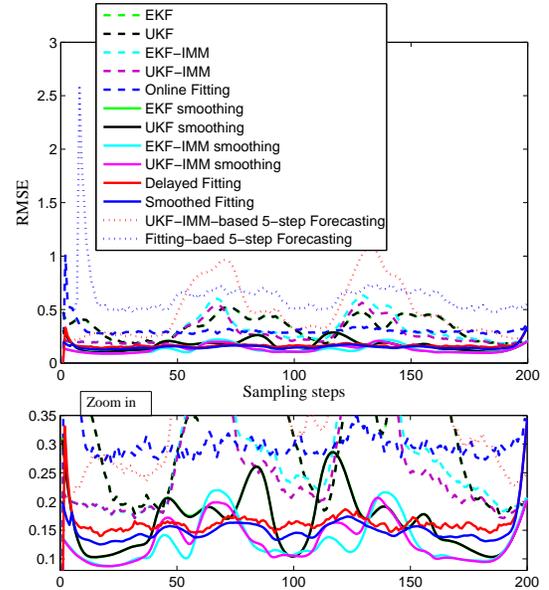}
\caption{RMSE of different nonlinear estimators over sampling steps when full and correct model information is provided to all estimators}
\end{figure}
\begin{figure*}\label{fig:4}
\centering
\includegraphics[width=12 cm]{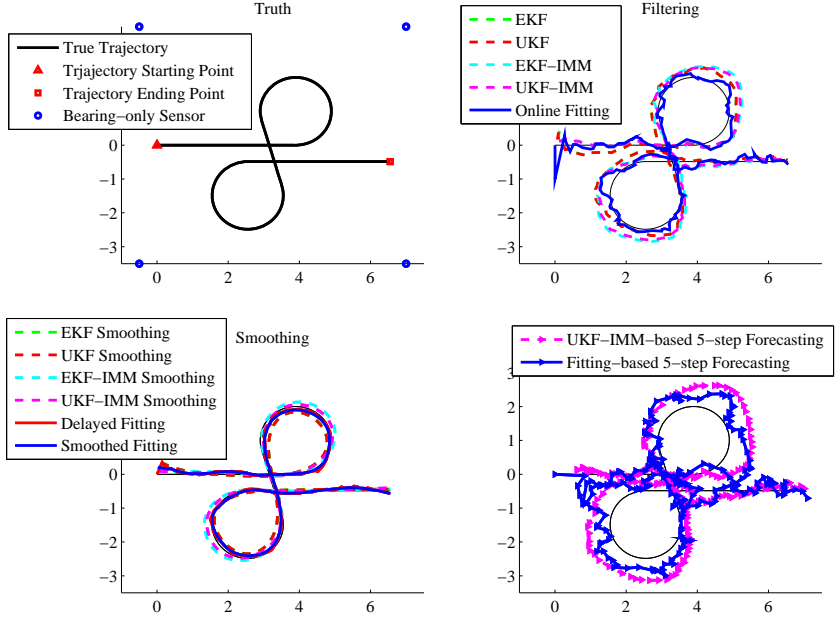}
\caption{Real trajectory and estimates given by different nonlinear estimators in one trial with incorrect a priori information about the sensors' noise}
\end{figure*}
\begin{figure}\label{fig:5}
\centering
\includegraphics[width=8 cm]{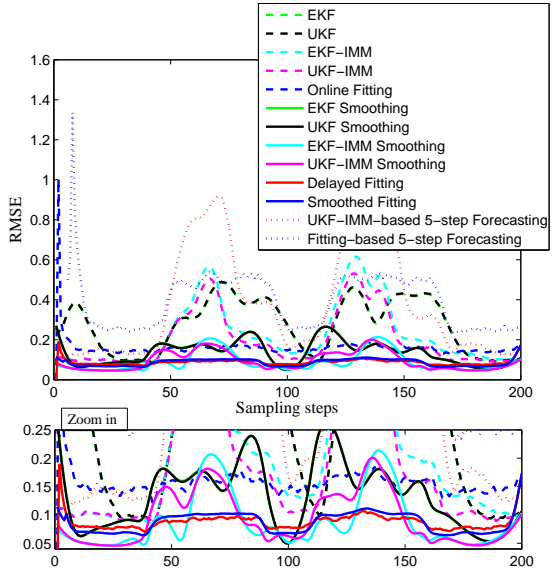}
\caption{RMSE of different nonlinear estimators over sampling steps when incorrect a priori information is provided about the sensors' uncertainty}
\end{figure}

Different from the previous example, at least two bearing sensors are needed to cooperate in order to project the bearing observations into the position space for O2 inference \cite{Li17clustering}. The nonlinear projection will lead to a bias if the bearing noise is not properly taken into account. For this reason, we perform the LS fitting directly on the bearing data with regard to the four sensors jointly, rather than on its projection in the position space. This releases the need of explicit statistical knowledge of the sensor observation noise. A sliding time window of 10 sampling steps (of length of totally $1$s) and a polynomial fitting fucntion of order $m=2$ were utilized. The fitting function is assumed as the same as \eqref{eq:21}. Given that the four sensors are of the same quality (while we do not really need to know $v_{k,i}$), the joint optimization function is given as
\begin{equation} \label{eq:26}
\underset{a_1,a_2,b_1,b_2}{\text{argmin}} \sum_{t=k_1}^{k_2} \sum_{i=1}^4 \bigg(\theta_{t,i}-\text{arctan}\Big(\frac{b_1+b_2t-s_{y,i}}{a_1+a_2t-s_{x,i}}\Big)\bigg)^2 ,
\end{equation}
where $k_2$ is the current time. 

Different to our previous simulation using simple linear measurement model that can be easily solved analytically, the above nonlinear formula is optimized by the LS curve fitting function: LSQCURVEFIT provided with the Optimization Toolbox of the Matlab software. All traditional estimators are initialized favorably with the true state $\mathbf{x}_0=[0,0,1,0,0]^T$ and covariance $\text{diag}([10.1,10.1,1.1,1.1,1]^T)$ and have the correct information about the sensor noise statistics. We initialize position estimates in our fitting approach at the first two sampling instants as $[0,0]^T$ and $[1,0]^T$, respectively. Then, the fitting (for smoothing, filtering and forecasting, respectively) is performed from the third sampling instant. This can be viewed as a \textit{hot-start} fitting as the information about the initial state $\mathbf{x}_0$ excluding covariance is used (while the fitting that uses no a priori information, as is done in the last simulation, is called \textit{cold-start}). 

We first set $\Sigma_v=0.01$. The simulation is performed with 100 Monte Carlo runs, each run consisting of 200 sampling steps (lasting 20 seconds) and using the same, deterministic trajectory but randomly generated observation series. The average performance of different estimators over 200 sampling steps is given in Fig.3. The RMSE over 200 sampling steps and the average running time per run are given in Table \ref{tab:2}. 

Regarding the estimation accuracy, the online fitting outperforms both the EKF and UKF using a single MPV model but slightly underperforms the EKF/UKF IMM. The smoothed fitting improves the delayed fitting while the latter is equivalent to EKS and UKS, all inferior to the EKF/UKF IMM smoothers. For 5-step ahead forecasting, our fitting approach outperforms the IMM approach. However, the computing speed of our fitting approaches is the fastest among all categories (whether for filtering, smoothing or forecasting) and is remarkably faster by using the LSQCURFIT tool as compared with the polynomial fitting used in the previous example. This is primarily because in our realization here, the coefficient parameters obtained in the last fitting process will be used as the initial parameters in the next fitting process, as in \eqref{eq:16}. This significantly reduces the optimization routine for solving \eqref{eq:26} and can be applied in most situations. 

Next, we change the simulation setup by reducing the real noise for all bearing sensor observation to $\Sigma_v=0.0025$ in \eqref{eq:25} without informing any estimators. All filters, smoothers and forecaster and the unbiased O2 inference still use the same information of $\Sigma_v=0.01$. The biased O2 inference and our fitting approaches that never use this information will not be affected. This situation corresponds to the realistic case that the user does not have correct information about the sensors' noise statistics. In this case, the performance for one run and the average performance over 100 runs for different estimators are given in Figs. 4 and 5, respectively. The RMSE and the average computing time per run are given in Table \ref{tab:3}. The results show that our fitting approaches benefit the most from the increased sensor accuracy. On the estimation RMSE, the online fitting outperforms the EKF/UKF and the EKF/UKF-IMM approaches. The smoothed and delayed fittings achieve the best performance on average, especially at the time periods when target maneuver occurs, such as $t=6-8$s and $t=13-15$s demonstrating better performance than the IMM approach that admits maneuver detection delay. It is interesting to note that there is a crossover of the performance of the delayed and the smoothed fitting, which slightly outperform each other at different model stages. As addressed, the latter may suffer somehow from overfit and therefore does not always perform better than the former. A further analysis about the overfit problem remains open. On the computing speed, our fitting approaches yield the fastest computing speed overall, demonstrating good real time running potentiality. 

The second simulation setup, in which complete and correct sensor noise information is unavailable, conforms more to reality than the first simulation setup where the model information, which is complete and correct, is simply too ideal. In the next example, we will demonstrate that even when perfectly modeled and parameterized, the suboptimal filters that are not so ideally initialized can perform worse than the data driven solutions that make little or no state model assumption. 

\begin{table}[h]
\caption{Average Performance of different nonlinear estimators (when completely correct sensor noise knowledge is used)}
\label{tab:2}
\begin{center}
\begin{tabular}{|c||c||c|}
\hline
\textbf{Estimators} & \textbf{Aver. RMSE} & \textbf{Compt. Time (s)}\\
\hline 
EKF (using WPV) & 0.3263	&0.0288\\
\hline
EKS (using WPV)	&0.1652	&0.0494\\
\hline
UKF (using WPV)	&0.3270	&0.0900\\
\hline
UKS (using WPV)	&0.1649	&0.1108\\
\hline
EKF-IMM	&0.2985	&0.1749\\
\hline
EKF-IMM smoother &0.1302	&1.2694\\
\hline
UKF-IMM	&0.2728	&0.9816\\
\hline
UKF-IMM smoother &0.1316 &2.7550\\
\hline
UKF-IMM 5-step forecaster &0.4845 &2.9821\\
\hline
Online Fitting	&0.3029	&0.0267\\
\hline
Delayed Fitting	&0.1631	&0.0446\\
\hline
Smoothed Fitting	&0.1500	&0.2623\\
\hline
Fitting-based 5-step forecaster	&0.6200	&0.0190\\
\hline
\end{tabular}
\end{center}
\end{table}

\begin{table}[h]
\caption{Average performance of different nonlinear estimators (when incorrect sensor noise knowledge is used)}
\label{tab:3}
\begin{center}
\begin{tabular}{|c||c||c|}
\hline
\textbf{Estimators} & \textbf{Aver. RMSE} & \textbf{Compt. Time (s)}\\
\hline 
EKF (using WPV) & 0.2716	&0.0404\\
\hline
EKS (using WPV)	&0.1341	&0.0587\\
\hline
UKF (using WPV)	&0.2725	&0.1557\\
\hline
UKS (using WPV)	&0.1338	&0.1736\\
\hline
EKF-IMM	&0.2247	&0.3078\\
\hline
EKF-IMM smoother &0.0883	&1.2402\\
\hline
UKF-IMM	&0.1965	&0.9463\\
\hline
UKF-IMM smoother &0.0939 &2.4897\\
\hline
UKF-IMM forecaster &0.3894 &2.6716\\
\hline
Online Fitting	&0.1599	&0.0248\\
\hline
Delayed Fitting	&0.0875	&0.0430\\
\hline
Smoothed Fitting	&0.0867	&0.3004\\
\hline
Fitting-based 5-step forecaster	&0.3937	&0.0148\\
\hline
\end{tabular}
\end{center}
\end{table}


\subsection{Ballistic target tracking}

\begin{figure}\label{fig:6}
\centering
\includegraphics[width=8 cm]{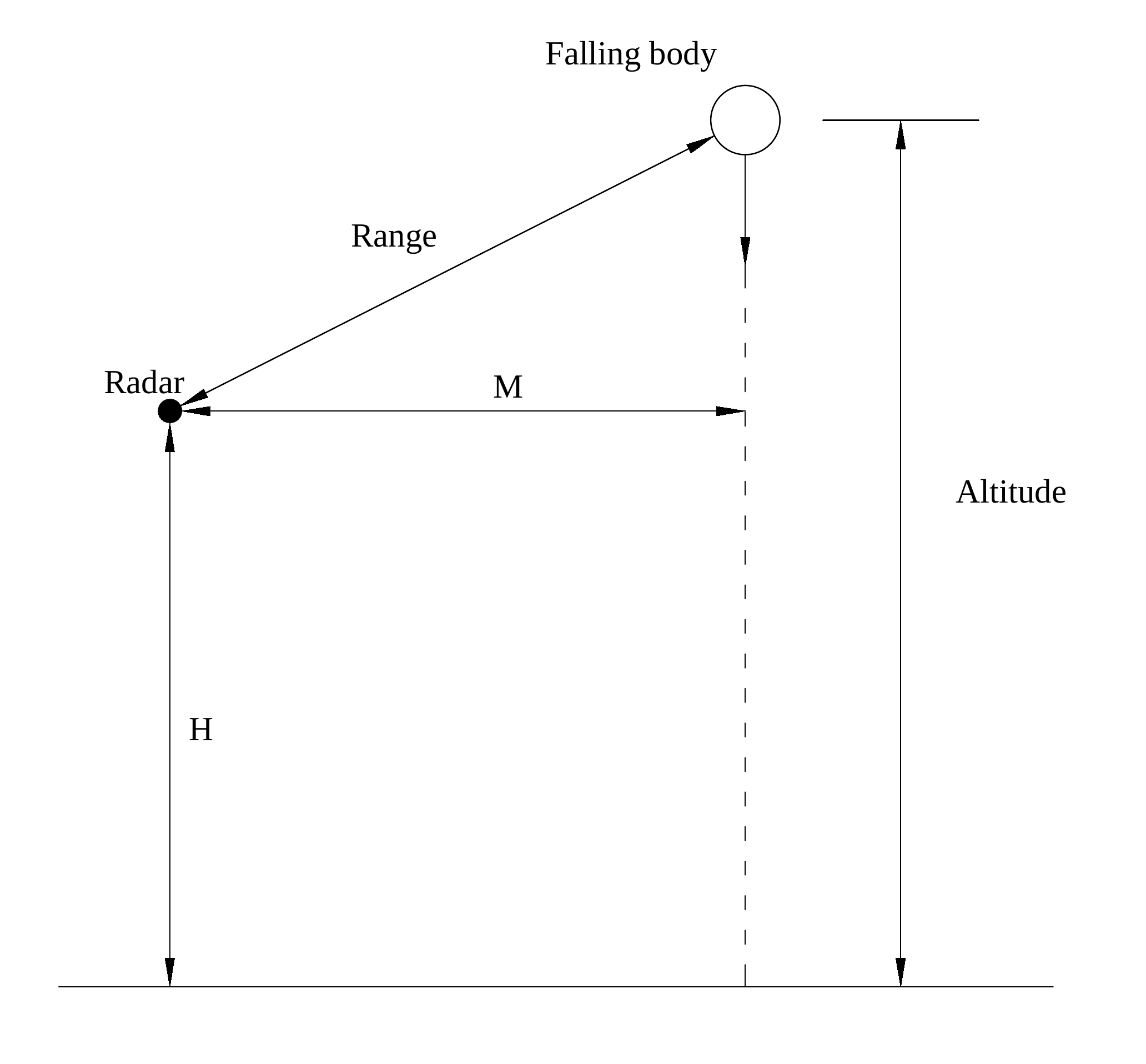}
\caption{Geometry of the vertically falling target}
\end{figure}
\begin{figure*}\label{fig:7}
\centering
\includegraphics[width=14 cm]{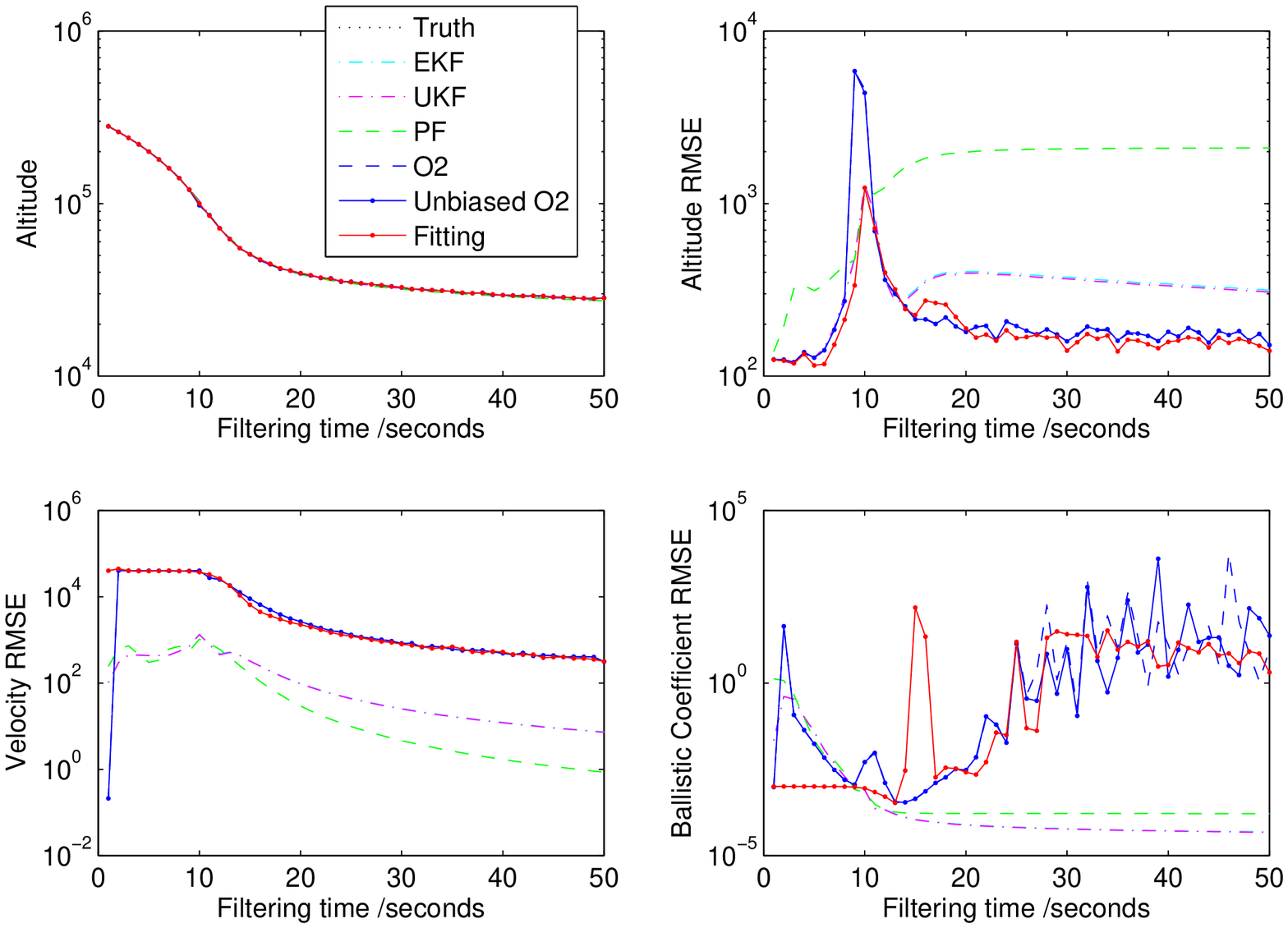}
\caption{Performances of different estimators when $R=10^4$: EKF, UKF, PF and the unbiased O2 inference are correctly modeled and provided with completely correct information of $R$ while the O2 inference and the fitting method does not need any information about $R$}
\end{figure*}
\begin{figure*}\label{fig:8}
\centering
\includegraphics[width=14 cm]{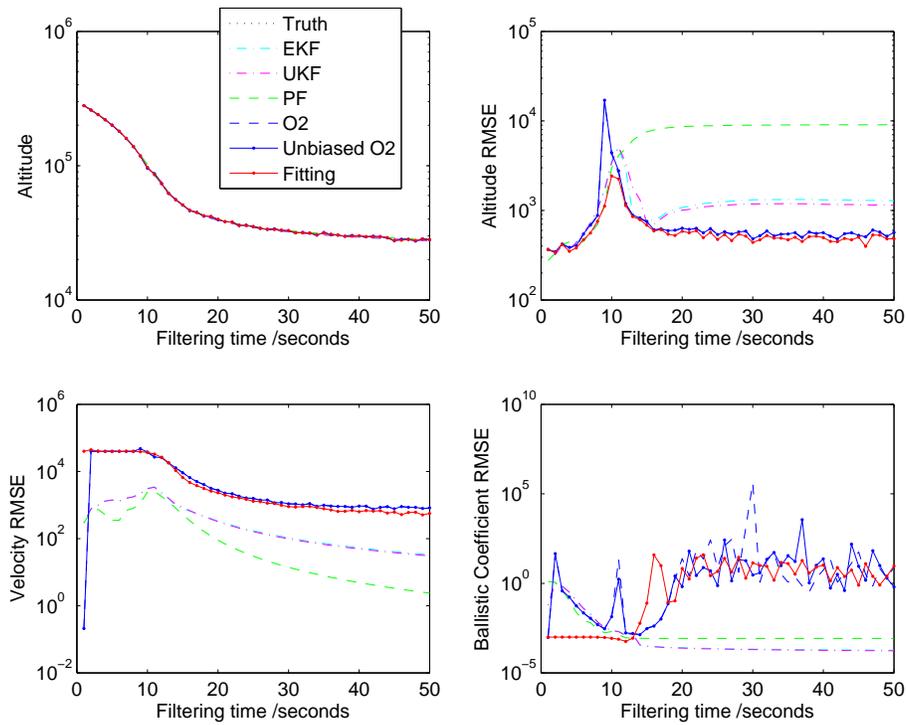}
\caption{Performances of different estimators when $R=10^5$: EKF, UKF and the unbiased O2 inference are correctly modeled and provided with completely correct information of $R$ while the O2 inference and the fitting method does not need any information about $R$}
\end{figure*}

In this example, the target is vertically falling, which is studied for testing nonlinear filters, such as \cite{Julier00} and \cite{Athans68}. The geometry of the scenario can be illustrated in Fig.6 with two known parameters: the horizontal distance of the radar to the target $M=10^5$ ft and the altitude of the radar $H=10^5$ ft.

The target state is modeled as $x_t = [h_t,s_t,c_t]^T$ consisting of altitude $h_t$, velocity $s_t$ and ballistic coefficient $c_t$. While only the altitude is the directly-observed state, we will also show how to infer the velocity and ballistic coefficient via fitting. The continuous-time nonlinear dynamics of the target is governed by the following differential equations 
\begin{equation} \label{eq:27}
\dot{h}_t=-s_t ,
\end{equation}
\begin{equation} \label{eq:28}
\dot{s}_t=-e^{-\gamma h_t}s_t^2c_t ,
\end{equation}
\begin{equation} \label{eq:29}
\dot{c}_t=0 ,
\end{equation}
where $\gamma=5\times10^{-5}$.

The discrete-time range observation is made every second and is given by
\begin{equation} \label{eq:30}
y_k=\sqrt[]{M^2+(h_k-H)^2} + v_k ,
\end{equation}
where the the observation noise is Gaussian $v_k\sim \mathcal{N}(0,R)$.

To simulate the ground truth, the initial state of the target is set as $\mathbf{x}_0 = [3\times10^5 \text{ft},2\times10^4 \text{ft/s},10^(-3) \text{ft}^{-1}]^T$. In accordance with \cite{Judd00} for high-precise approximation of the differential equations, a fourth order Runge-Kutta method based on 64 iterations every second between two successive observations is employed for simulating the deterministic ground truth. The fourth order Runge-Kutta method is also used in the estimators for accurate simulation of the state motion. 

In this example, the online fitting to compare with the O2 inferences and three typical nonlinear filters including EKF \cite{Athans68}, UKF \cite{Julier00} and PF. To initialize them for tracking, a priori information of the initial state of the target is given as $\mathbf{x}_0 =[3\times10^5 \text{ft},2\times10^4 \text{ft/s},3\times 10^(-3) \text{ft}^{-1}]^T$ with error covariance $\mathbf{P}_0=[10^6 \text{ft},4\times10^6 \text{ft/s},10^(-4) \text{ft}^{-1}]^T$. This is the same as that in \cite{Julier00, Athans68}. That is, a priori knowledge about the target altitude and velocity is ideally consistent with the truth but knowledge regarding the ballistic coefficient is bad. In particular, the Gaussian observation noise is of relatively small variance, which will yield a steep Gaussian likelihood function. As a result, the standard sampling importance resampling (SIR) PF will suffer from the sample degeneracy/impoverishment significantly\cite{Li15resampling}
. To combat this, we use a likelihood function with heavy tails as follows
\begin{equation} \label{eq:31}
p(y_k|h_k^{(i)}) \propto \text{exp}\bigg(-\frac{(y_k-y_k^{(i)})^2}{\underset{i}{\text{max}}\{(y_k-y_k^{(i)})^2\}}\bigg),
\end{equation}
where $y_k^{(i)}=\sqrt[]{M^2+(h_k^{(i)}-H)^2}$ and $h_k^{(i)}$ is the estimated altitude of the $i$th particle of the total $200$.

Given the latent constraint that: the target is falling vertically and will not make any movement in the horizontal direction, the O2 inference estimates the altitude $h_k$ based on the triangulation between $h_k$, $M$ and $y_k$ as follows:
\begin{equation} \label{eq:32}
\hat{h}_k=+/-\sqrt[]{y_k^2-M^2} + H ,
\end{equation}
where the sign will change from positive to negative (only once) at an altitude about $h_k \approx H$ during the entire tracking process. 

More specifically, the sign can be determined based on the elevation angle of the radar in practice (if available) or based on another latent rule that: \textit{the ball is falling in a single direction and with a velocity that accelerates with time}. That is, we have the contextual information of ``\textit{accelerated falling}" which forms a constraint on the altitude and velocity as follows:
\begin{equation} \label{eq:33}
\left\{
\begin{array}{ll} 
\hat{h}_k<\hat{h}_{k-1}\\
\hat{s}_{k-1} < \hat{s}_k
\end{array} \right. .
\end{equation}

When the statistics of the sensor observation noise is known, de-biasing shall be applied for which we applied the Monte Carlo de-biasing approach \cite{Li16O2, Li17clustering} using 100 samples. This is referred to unbiased O2 inference while that given by \eqref{eq:32} without debiasing is referred to as biased O2 inference.

As addressed earlier, we can assume a reasonable trajectory function on both altitude and velocity to carry out fitting for estimation. Here, the directly-observed state is the altitude and its fitting estimation shall rely only on the observation, while for velocity, the deterministic dynamic model \eqref{eq:27}-\eqref{eq:29} can be used. For both altitude and velocity, we use a sliding time window of no more than 5 sampling steps (as we found that both trajectories are very smooth) and 3-order fitting function. That is, the altitude function is given as
\begin{equation} \label{eq:34}
h_t=a_1+a_2t+a_3t^2 ,
\end{equation}
and the corresponding object minimizing function
\begin{equation} \label{eq:35}
\underset{a_1,a_2,a_3}{\text{argmin}} \sum_{t=k_1}^{k_2} \big( y_k - \sqrt[]{M^2+(a_1+a_2t+a_3t^2)^2}\big)^2 .
\end{equation}

Once the altitude trajectory function \eqref{eq:34} is obtained, its derivation gives that
\begin{equation} \label{eq:36}
\dot{h}_t = a_2+2a_3t ,
\end{equation}

Similarly, the velocity function can be assumed
\begin{equation} \label{eq:37}
s_t=b_1+b_2t+b_3t^2 ,
\end{equation}
and consequently, we have its derivation 
\begin{equation} \label{eq:38}
\dot{s}_t = b_2+2b_3t .
\end{equation}

The velocity function should be determined such that the discrepancies between both sides of \eqref{eq:27} and \eqref{eq:28} are minimized. By comparing \eqref{eq:36} with \eqref{eq:27} and \eqref{eq:38} with \eqref{eq:28}, these two discrepancies can be written as, respectively
\begin{equation} \label{eq:39}
\Phi_1:= \dot{h}_t -(-s_t) ,
\end{equation}
\begin{equation} \label{eq:40}
\Phi_2:= \dot{s}_t- (-e^{-\gamma h_t}s_t^2\hat{c}_t) ,
\end{equation}
where 
$\hat{c}_t$ is calculated using the data from the previous sampling time instant, based on \eqref{eq:29}, as follows
\begin{equation} \label{eq:41}
\hat{c}_t :=-\frac{\dot{s}_{t-1}}{e^{-\gamma h_{t-1}} s_{t-1}^2}.
\end{equation}
We are able to do so because the ballistic coefficient is a constant (such is known a priori). However, an initial estimate of it is needed in the first round of fitting, for which we assume the same as that in the filters, i.e., $c_0=3\times 10^{-3} \text{ft}^{-1}$. 

Given that \eqref{eq:39} and \eqref{eq:40} are equally weighted, the joint optimization function in the LS manner can be written as
\begin{equation} \label{eq:42}
\underset{b_1,b_2,b_3}{\text{argmin}} \sum_{t=k_1}^{k_2} (\Phi_1^2+\Phi_2^2).
\end{equation}

First, we set sensor noise the same as that in \cite{Julier00} by using $R=10^4$. This information is precisely provided to the EKF, UKF, PF and the unbiased O2 inference while the biased O2 inference does not need it. The simulation results are given in Fig. 7 for the altitude truth and estimates given by different estimators in one trial and the RMSEs of the altitude, velocity and ballistic coefficient, respectively. It is shown that at altitudes near $h_t \approx H$, all estimators are highly inaccurate. Surprisingly, except near $h_t \approx H$ when the calculation given by \eqref{eq:32} is very inaccurate, the O2 inference outperforms the EKF/UKF/PF remarkably. This indicates that the filters are in fact ineffective most of the time according to the definition of ``\textit{effectiveness}" given in \cite{Li16O2}. However, both de-biasing and fitting approaches have not improved the biased O2 inference by much. The average altitude RMSEs and computing times of different estimators are given in Table \ref{tab:4}. Overall, the online fitting inference achieves the best performance while the PF does the worst in altitude estimation. Both the (biased and unbiased) O2 inferences and the online fitting approach are inferior to the ideally-modeled filters in the estimation of the velocity and constant ballistic coefficient, which are not directly-observed variables. On the computing speed, the O2 inference is unsurprisingly much faster than the others while the PF is the slowest. 

Second, we apply a larger sensor noise $R=10^5$ which is correctly provided to all estimators and all the other settings remain the same. The simulation results are given in Fig. 8 and Table \ref{tab:5} for similar contexts to that of Fig. 7 and Table \ref{tab:4}, respectively. Most of the trends in Fig. 8 are similar to those in Fig. 7. We can see that the O2 inference, unbiased O2 inference and our fitting approach outperform all filters including the EKF, UKF and the PF for altitude estimation. While the de-biasing still does not improve the O2 inference, the fitting does so
. For more insights, Figs. 7 and 8 show that the fitting approach benefits the most around time 10 (when $h_t \approx H$) when all estimators suffer from unstable estimation. The biased and unbiased O2 inferences and the online fitting approach are again significantly inferior to filters in estimating the velocity and ballistic coefficient. This exposes one limitation of the data driven approaches including the O2 inferences and the online fitting approach, to which we need to develop more thorough data mining solutions to explore the deterministic model information hidden in \eqref{eq:27}-\eqref{eq:29}. At the current stage, we primarily concentrated on the directly-observed state and position-trajectory inference.

\begin{table}[h]
\caption{Average altitude RMSE and computing time of different estimators ($R=10^4$)}
\label{tab:4}
\begin{center}
\begin{tabular}{|c||c||c|}
\hline
\textbf{Estimators} & \textbf{Aver. Altitude RMSE(ft)} & \textbf{Compt. Time (s)}\\
\hline 
EKF& 355	&0.1517\\
\hline
UKF	&349	&0.1588\\
\hline
PF &1664	&24.63\\
\hline
Biased O2 &396	&2.7$\times 10^{-4}$\\
\hline
Unbiased O2	&392	&0.0022\\
\hline
Fitting	&212	&2.324\\
\hline
\end{tabular}
\end{center}
\end{table}

\begin{table}[h]
\caption{Average altitude RMSE and computing time of different estimators ($R=10^5$)}
\label{tab:5}
\begin{center}
\begin{tabular}{|c||c||c|}
\hline
\textbf{Estimators} & \textbf{Aver. Altitude RMSE(ft)} & \textbf{Compt. Time (s)}\\
\hline 
EKF& 1262	&0.1518\\
\hline
UKF	&1213	&0.1601\\
\hline
PF &6970	&16.41\\
\hline
Biased O2 &1002	&3.18$\times10^{-4}$\\
\hline
Unbiased O2	&1031	&0.0023\\
\hline
Fitting	&613	&2.517\\
\hline
\end{tabular}
\end{center}
\end{table}

\section{Conclusion}  \label{conclusion}

For a class of target tracking problems with poor a priori knowledge about the system, we presented an online sensor data fitting framework for approximating the continuous-time trajectory function. This leads to a unified methodology for joint smoothing, tracking and forecasting and provides a flexible and reliable solution to use information such as ``\textit{the target is descending}", ``\textit{the target is about passing by a location}" or ``\textit{the target goes to a fixed destination}". Such information is common and important in reality, but is often overlooked or treated in an ad-hoc manner in existing solutions because they cannot be quantitatively defined as statistical knowledge. 

In a variety of representative scenarios, the proposed methods perform comparably to classic suboptimal algorithms that have complete and correct model information and can even outperform them if, more commonly in reality, they are provided with incorrect sensor statistical information or are improperly initialized. Moreover, the present sliding time window fitting approach does not need to take ad-hoc multiple/adaptive models to handle target maneuver, as long as the target trajectory remains smooth over the time window. This adds greatly to the reliability, flexibility and ease of implementation of the framework. 

Unifying the tasks of smoothing, tracking and forecasting on a single estimation framework is of high interest to many real world problems of significance, where the history, the current and the future of the target's state are desired simultaneously
. Also, continuous time trajectory acquisition is essential for detecting and resolving potential trajectory conflicts when multiple targets exist or compete and for analyzing/learning the target movement pattern. All of these, together with the rapid escalation and massive deployment of massive sensors, will make sensor data-learning/fitting approach more promising. 

There is broad space for further development in this direction, including, better data mining solutions to infer the indirectly-observed variables from the directly-observed variables (especially when the observation is sparse)
, to obtain further statistical knowledge of the estimate (e.g., the accuracy given in the manner of variance) and data-to-trajectory association for handling multiple (potentially interacting) targets. 

\section*{Appendix}

\subsection {Lagrange remainder for linearization}
We analyze the linearization error caused by converting a nonlinear fitting function to a linear one 
based on the Taylor series expansion. 
A Taylor series of a real function $f(\mathbf{x})$ about a point $\mathbf{x}=\mathbf{x}_0$ is given by
\begin{equation*}
f(\mathbf{x})=f(\mathbf{x}_0)+\dot{f}(\mathbf{x}_0)(\mathbf{x}-\mathbf{x}_0)+\cdots+\frac{1}{n!}f^{(n)}(\mathbf{x}_0)(\mathbf{x}-\mathbf{x}_0)^n+R_n ,
\end{equation*}
where $R_n$ is a remainder term known as the Lagrange remainder (also known as truncation error), which is given by
\begin{equation*}
R_n=\frac{f^{(n+1)}(\bar{\mathbf{x}})}{(n+1)!}(\mathbf{x}-\mathbf{x}_0)^{n+1} ,
\end{equation*}
where $\bar{\mathbf{x}} \in [\mathbf{x}_0,\mathbf{x}]$ lies somewhere in the interval $[\mathbf{x}_0,\mathbf{x}]$.  

This indicates that the closer the prediction data x is with $\mathbf{x}_0$, the 
smaller $R_n$. This explains why piecewise/sliding time-window fitting, is highly suggested in our approach. 

\subsection{Recursive Least Square Adaptive Filter}
Let $y_k= x_k^Tc_k+e_k$ be a general 1-dimensional linear fitting model where $x_k$ and $y_k$ are the system input (regressor) and
output at time $k$, $e_k$ represents an additive noise (i.e., the fitting error to be admitted) and $c_k$ is the parameter to be recursively estimated for minimizing the square error $\sum_{t=1}^{k}\lambda^{k-t} (y_t-x_t^Tc_t)^2 $. The recursive LS algorithm updates the parameter estimate by the recursion in a time-window
\begin{equation*}
\hat{c}_k = \hat{c}_{k-1} + \mathcal{G}_k(y_k-x_k^Tc_{k-1}),
\end{equation*}
\begin{equation*}
\mathcal{G}_k = \frac{P_{k-1}x_k}{\lambda + x_k^TP_{k-1}x_k} ,
\end{equation*}
\begin{equation*}
P_k = \frac{1}{\lambda}\bigg( P_{k-1}-\frac{P_{k-1}x_kx_k^TP_{k-1}}{\lambda + x_k^TP_{k-1}x_k} \bigg) ,
\end{equation*}
where the forgetting factor $\lambda$ is usually chosen in $[0.9, 0.999]$ to reduce the influence of past data or chosen as $1$ to equally treat all data in the time window, and the matrix $P_t$ is related to the covariance matrix, but $P_t \neq \text{Cov}(\hat{C}_k)$.



%



\ifCLASSOPTIONcaptionsoff
  \newpage
\fi



\bibliographystyle{IEEEtran}
\bibliography{IEEE_STF}
%



%




\end{document}